\documentclass[12pt]{article}
\usepackage{amsmath}
\usepackage{amssymb}

\renewcommand{\abstractname}{\underline{\rm A b s t r a c t}}
\begin{document}

\smallskip
\begin{center}                                
    CALCULATIONS OF MESON MASS SPECTRA IN THE MODEL OF QUASI-
                       INDEPENDENT QUARKS {\footnote {The lecture 
at the XLI PNPI Winter School, February 19-24, 2007, St.-Petersburg, Repino}}

\medskip

                         V.V. Khruschov

RRC "Kurchatov Institute", Moscow, 123182, Ac. Kurchatov Sq. 1 
\end{center}                                
                            
\begin{abstract}                                
   A  method  of  calculations  of meson  mass  spectra  in  the
framework of the model of quasi-independent quarks is discussed.
Meson mass spectra evaluated with the help of the Dirac equation
with   the   quasi-Coulombic    and      confinement
potentials,  as  well  as with the help of phenomenological  mass
formulae,  are  presented. Parameters of  the  quasi-independent
quarks  model, which are calculated on the basis of mass  values
of pseudoscalar and vector mesons, are given. Possible relations
between  parameters of the quasi-independent  quarks  model  and
constants  of  the  Standard  model of  strong  and  electroweak
interactions are considered.                          
\end{abstract}

\medskip 
\begin{center}
 1. \underline{Introduction} 
\end{center}
\smallskip

It is well known that the calculation of hadron mass spectra on the level 
 of experimental data precision \cite{eidel}  still remains among the unsolved
QCD problems due to some technical and conceptual difficulties, which are
related mainly with the nonperturbative effects, such as confinement and
spontaneous chiral symmetry breaking. For instance, although recently a
considerable progress in lattice QCD was achieved, at
 present the calculations of hadron mass values on the basis of the first
principles of QCD  demand too much
computation time especially for the light hadrons. So data
interpretations and  calculations 
of hadron characteristics are frequently
carried out with the help of  phenomenological  models. Moreover
 the identification of hadrons bearing
constituents, distinguishable from standard ones, now is the topical problem
for strong interaction physics. For this purpose a detailed
description of characteristics of standard hadrons is needed.
 Especially it concerns the
evaluation of  masses for ground and exited hadron states with
light quarks and/or antiquarks. 

 One of the most interesting and simple hadron models is the
relativistic model for quasi-independent quarks. The main
statement of this model is that hadron's properties can be
described considering the hadron as a system of independent
constituents (or quasi-independent ones with weak residual
interactions), which move in some mean self-consistent field \cite{bog}.
For the description of constituent's motion the one-particle
equation (Schrodinger, Dirac or Klein-Gordon-Fock equation) can be
used. The confinement of color particle is available by using of a
linear rising potential, for which the hypothesis of its flavour
independence has been proposed \cite{doro,mart,quroth}.
  The relativistic model for quasi-independent quarks has been
applied with the particular potential entered in the Dirac
equation for the description of meson properties in Refs. \cite{khrusas,ks}.
The potential used is a generalization of Cornell potential and
consist of a relativistic vector quasi-Coulomb potential and a
relativistic scalar linear rising potential. In the framework of
this model the hypothesis of universality of confinement potential
offered in Refs.\cite{doro,mart,quroth} has been confirmed for heavy and light
quarks. The coefficient of linear rising scalar potential (string
tension) has been found to be $\sigma = 0.20\pm 0.01 GeV^2$.
Below we present the results of  calculations of mass spectra of  mesons
and model parameters, which may be related directly to constants of 
Standard Model.  In order to obtain the consistent treatment of $1^{--}$
and $0^{-+}$ mesons it is nessasary to take into account spin-spin
interaction between quark and antiquark as well as the values of the meson
mean field energy \cite{ks}. Note that the account of these terms are most important
in the case of the pseudoscalar mesons. We made a link of quasi-independent
quark model with the constituent quark model and propose that 
the values of the meson mean field energy belong equidistant energy levels.
We present also mass
spectra of excited meson states\ with $I=1$ consisting of $u-$, $d-$quarks
and antiquarks with help of analytical mass formulae, which have the structure
peculiar for mass formulae of the independent quark model. These
formulae may be considered as the generalisations of the Chew-Frautschi and
Nambu-Veneziano relations for the Regge trajectories of superlight mesons
(mesons which consist from $u-$, $d-$quarks and antiquarks only). We
compare results obtained with existing data and make predictions for mass
values of unobserved meson states. Besides
that, we consider  possible  structures of some mesons which are
under discussion at present, such as scalar mesons and vector excited mesons.

The report is organized as follows. We briefly review the basic equations
and the main statements of the model in Section 2. Here the Dirac equation
in the external potential with the Lorentz scalar and vector
parts is transformed into a form suitable for  numerical
calculations. The values of model parameters 
and meson masses are presented in Section 3. In Section 4 
the phenomenological mass formulae for isovector 
radially and orbitally exited $\overline{q}q^{\prime }-$mesons
are presented and the the results of calculations for mass
spectra of  exited mesons are displayed. In Section 5 the model
of quasi-independent quarks is related to the constituent
quarks model. We discuss the results obtained and the problems of 
identification of some meson states in vector and scalar channels 
in the last section.

\begin{center}
 2.\underline{ Basic statements of the relativistic
independent quark model}
\end{center}
\smallskip

According to the main statement of the independent-quark model a hadron is
considered as a system composed of a few non-interacting with each other
directly, valency constituents (quarks, 
diquarks and constituent gluons) having the coordinates $\mathbf{r}_{i}$, 
$i=1,...,N$, and moving in some mean field. We suppose that this field is a
colour singlet confining field which is produced by the constituents 
and nonperturbative QCD vacuum and
takes into account the effects of creation and annihilation of a sea of 
$q\bar{q}$ pairs as well. Furthermore, to simplify calculations it is assumed
that this mean field is spherically symmetric and is 
a quasi-classical object possessing some energy $E_{0}$ inside 
a hadron. Each of $N$ constituents interacting
with the spherical mean field gets the state with a definite value of its
energy $E_{i}$, so that the hadron mass can be evaluated as
\begin{equation}
M_h=E_0+E _1+...+E _N.
\end{equation}
The wave function $\psi_i$  
for any constituent is a solution of a single-particle equation with the mean
field static potential $U(\mathbf{r}_{i})$.
Angular dependence of the single-particle wave functions in a stationary 
state for the spherically symmetric potential may be separated in a well-known
manner, and in order to evaluate the constituent energies $E_i$ it is
necessary to solve the radial equations with the model potentials for each
constituent. Note that it is impossible to evaluate $E_{0}$ without
additional assumptions, and this quantity take as a phenomenological 
parameter in the framework of the model.

According to these assumptions the mass $M_{m}$ of the $\bar{q}q$ -
meson in the main approximation can be evaluated as 
\begin{equation}
M_{m}=E_{0m}+E_{1}(n_{1}^{r},j_{1})+E_{2}(n_{2}^{r},j_{2}).
\label{mava}
\end{equation}

\noindent where $E_{i}(n_{i}^{r},j_{i})$ , $i=1,2$, are the energy spectral functions
or the mass terms for the $i-$th quark (antiquark) and represents the
relativistic effective energy of the $i-$th quark/antiquark moving in the
mean field inside the meson. Here $n_{i}^{r}$ and $j_{i}$ are the radial
quantum number and the quantum number of the angular moment correspondingly
for the $i-$th constituent.

 The term $E_{0m}$
contains the contribution from the mean field energy and the possible
nonpotential corrections, which cannot be evaluated in the frame of
mean field approximation. This term has a nonzero value only for some 
meson ground  states. 
 The terms $E_{i}(n_{i}^{r},j_{i})$, $i=1,2$, which represent the energies of
the constituents in the mean field, should be evaluated for quarks from the solution
of the Dirac equation (for diquarks and constituent gluons from the solutions
of Klein-Gordon-Fock equation):
\begin{equation}
\sqrt{\lambda _{i}+m_{i}^{2}}\psi _{i}(\mathbf{r_{i}})=\left[ (%
\mbox{\boldmath
$\alpha_i$}\mathbf{p_{i}})+\beta _{i}(m_{i}+V_{0})+V_{1}\right] \psi _{i}(%
\mathbf{r_{i}}),
\label{diq}
\end{equation}
with $E_{i}(n_{i}^{r},j_{i})=$ $\sqrt{\lambda _{i}+m_{i}^{2}}$,
 $i=1,2$,  $V_{0}(r)=$ $%
\sigma r/2$ and $V_{1}(r)=$ $-2\alpha _{s}/3r$, where the model parameters $%
\sigma $ and $\alpha _{s}$ have meanings of the string tension and the
strong coupling constant at small distances, correspondingly.
As it is seen from  Eq.(\ref{diq}),
 the addition of some constant to the scalar linear confining potential $%
V_{0}(r)$ is equivalent to the addition of the same constant to the quark
mass and \textit{vice versa}, while the addition of some constant to the
vector quasi-Coulombic potential $V_{1}(r)$ is equivalent to the energy
shift of opposite sign.

It is well-known that the solutions of Eq.(\ref{diq}) with the total angular
momentum $j$ and its projection $m$ can be represented as 
\begin{equation*}
\psi (\mathbf{r})\propto \left( 
\begin{array}{r}
f(r)\Omega _{jl}^{m}(\mathbf{n}) \\ 
-ig(r)(\mbox{\boldmath $\sigma$}\mathbf{n})\Omega _{jl}^{m}(\mathbf{n})%
\end{array}
\right) ,
\end{equation*}
where $\mathbf{n}=\mathbf{r}/r$, the subscript $i$ here and below is
omitted. If $k=-\omega (j+1/2)$, where $\omega $ is an eigenvalue of the
space-parity operator, the system of the radial Dirac equations for the
fermion in the mean field with the definite energy sign and spin projection
reads
\begin{eqnarray}
(\sqrt{\lambda+m^2}-V_0-V_1-m)f = -\frac{(1-k)}{r}g-g^{\prime},  \notag \\
(\sqrt{\lambda+m^2}+V_0-V_1+m)g =\quad\frac{(1+k)}{r}f+f^{\prime}.  \label{se}
\end{eqnarray}

Using Eqs.(\ref{se}) one can derive the second order equation for the "large"
component $f(r)$, and then making a substitution
\begin{equation}
\varphi(r)=rf(r)\left[ V_0(r)-V_1(r)+m+\sqrt{\lambda+m^2}\right]^{-1/2}
\nonumber
\end{equation}
one comes on to the model radial equation for $\varphi(r)$ in the following
form:
\begin{equation*}
\varphi ^{^{\prime \prime }}+\lambda \varphi =[
V_0^2-V_1^2
+2(mV_0 + \sqrt{\lambda +m^2}V_1)+\frac{k(k+1)}{r^2}+\frac{3(V_0^{^{\prime }}
-V_1^{^{\prime }})^2}{4(\sqrt{\lambda +m^2}-V_1+V_0+m)^2}
\end{equation*}
\begin{equation}
+\frac{k(V_0^{^{\prime }}-V_1^{^{\prime }})}{r(\sqrt{\lambda +m^2}-V_1+V_0+m)%
}-\frac{(V_0^{^{\prime \prime }}-V_1^{^{\prime \prime }})}{2(\sqrt{\lambda
+m^2}-V_1+V_0+m)}{\Huge ]}\varphi 
\label{gre}
\end{equation}

The ground states of standard $\bar qq'-$mesons consist of the 
S-wave $1^{--}$ and $0^{-+}$ mesons  and the most extensive set of
precise data \cite{eidel} for heavy mesons is related to the radially exited
S-wave $1^{--}$ mesons. For this type of mesons $k=1$ and the parameter 
$\lambda $ entering Eq.(\ref{gre}) can be calculated with the help of the 
radial equation:
\begin{equation*}
\varphi ^{^{\prime \prime }}+\lambda \varphi =\{-\frac{4\alpha _s\sqrt{%
\lambda +m^2}}{3r}-\left( \frac{2\alpha _s}{3r}\right) ^2+m\sigma r+
\end{equation*}
\begin{equation}
+\left( \frac{\sigma r}2\right) ^2-\left[ \frac{\sigma r}2\left( m+\sqrt{%
\lambda +m^2}\right) +\left( \frac{\sigma r}4\right) ^2+\frac{5\alpha
_s\sigma }6-\frac{\alpha _s^2}{3r^2}\right] {\Huge /}
\label{effp}
\end{equation}
\begin{equation*}
\left[ \frac{2\alpha _s}3+r\left( m+\sqrt{\lambda +m^2}\right) +\frac{\sigma
r^2}2\right] ^2\}\varphi.
\end{equation*}

The right-hand side of Eq.(\ref{effp}) has a singularity at the origin, and when $%
r\to 0$ it behaves as $3/4r^2-4\alpha _s^2/9r^2$. Therefore one should keep $%
\alpha _s<3/2$ in order to prevent a fall at the origin. At the infinity $%
r\rightarrow \infty $ the effective potential behaves as the oscillatory one
and tends to $\sigma ^2r^2/4$. The form (\ref{effp}) of the effective potential
partly clarify some varieties of the confining potential used in potential 
models. We see, that for light quarks the confining potential is mainly 
oscillatory, while for heavy quarks it is  linearly rising
in the main approximation.

\begin{center}
3. \underline{Evaluations of model parameters and meson masses} 
\end{center}
\smallskip

The model equation (\ref{effp}) can be solved only by numerical methods. For
calculating its eigenvalues the computer code algorithm, which was based on
the Numerov three point recurrent relation \cite{num}, has been used jointly with
the regularization procedure for the singular potential at $r=0$, which is
analogous to the procedure worked out, for instance, in Ref.\cite{buen}. 
 The accuracy criterion for fitted parameters is the maximum
value of the  errors for evaluated hadron masses, which should not exceed
 typical experimental errors 
$30\div 40$ MeV. Thus, the parameter errors written below 
are our estimation of systematic errors of the model,
which have been found in this manner using
mesons mass spectra data \cite{eidel}. 

The brief description of parameter fitting is as follows.  When calculating the values of the model parameters we do not suppose 
\textit{a priori} the validity of the hypothesis of flavor independence for
the confining potential. First of all the model parameters $m$, $\alpha _s$
and $\sigma $ for the $b\bar b$ and $c\bar c$ radial excitation of $1^{--}$
states were found. The fits were carried out for each meson family
independently. The existing experimental data allowed to find these
parameters as well as to prove the validity of the model. In this manner it
was found that the value of $\sigma $ is the same for the $b\bar b$ and $%
c\bar c$ states within the systematic errors of the model and equal to 
$\sigma =(0.20\pm 0.01) GeV^2$.

Taking into account that there are no well established data for the higher
radial excitations of $1^{--}$ light mesons, the $\sigma $ value obtained
was used for calculating $m$ and $\alpha _s$ for light quarks. Note that
when doing the calculations we neglect the isotopic mass splitting of the $u$%
- and $d$-quarks because it is beyond of the model accuracy. 
It should be noted that the terms $E_{i}(n_{i}^{r},j_{i})$, $i=1,2$, and $%
E_{0m}$ in Eq.(\ref{mava}) represent the main contributions to the energy of the
system due to the interaction of the constituents with the mean field and
the mean field energy, respectively. However, it is obvious, that 
in the spin independent potential the
constituent energies $E_{i}(n_{i}^{r},1/2)$ are the same for the $1^{--}$
and $0^{-+}$ mesons with the identical radial quantum numbers and the mass
differences can be the result of the spin-spin interaction and the different
values of the $E_{0m}$ terms. 

In order to estimate the spin-spin interaction between quark and antiquark
we take into account the following expression for the interaction of this
type which has been used in the relativized quark model \cite{godf}:
\begin{equation}
V_{s_1s_2}=\frac{32\pi\alpha_s \mbox{\boldmath $s_1$}\mbox{\boldmath $s_2$} 
}{9E_1E_2} \delta^3(\mathbf{r}), 
\end{equation}
\smallskip 
\noindent where $\mbox{\boldmath $s_1$}$ and $\mbox{\boldmath $s_2$}$ 
are the spin operators and $E_1$ and $E_2$ are the energies of quark
and antiquark, respectively. Evaluation of the expectation value of this
operator in the first order of perturbation theory for the wave functions of
quark and antiquark shows that the spin-spin interaction depends mainly 
on the $E_1$ and $E_2$,  the total spin of
quark-antiquark system and the strong coupling constant $\alpha_s$. So one
can use the modified mass formula for S-wave $1^{--}$ and $0^{-+}$ mesons,
which takes into account spin-spin interaction between quark and aniquark,
in the following form \cite{ks}:
\begin{equation}
M_{m}=E_{0m}+E_{1}(n_{1}^{r},1/2)+E_{2}(n_{2}^{r},1/2)+4<%
\mbox{\boldmath
$s_1$}\mbox{\boldmath $s_2$}>_{q_{1}\bar{q}_{2}}V_{ss}
\label{me}
\end{equation}

Although in some cases for vector heavy mesons the
contributions of spin-spin interaction are less than the model precision,
nevertheless, for pseudoscalar mesons for 
all bound states of quark and
antiquark, the spin-spin contributions should be taken into account. 
The magnitude of the spin-spin interaction constant $V_{ss}$ is 
of the order of  $100$ MeV for light quarks and of the order of  $10$ MeV 
for heavy quarks, while the values of  the $E_{0m}$ parameter are:
$
 E_0(K)=-118 MeV
$, $
E_0(\eta _c)= -230 MeV
$, $
E_0(J/\Psi)=-125 MeV 
$, $
E_0(\pi ) = -236  Mev
$, $
E_0(\eta_b) = -600 MeV
$, $
E_0(\Upsilon) =-450 MeV
$, $
 E_0(D_s)=-112 MeV
$, $
 E_0(B_c)=-270 MeV
$. The values of $E_{0m}$ for mesons, which were not pointed out above, 
within the model accuracy may be set to zero. 

We obtain the following values for the coupling $\alpha _s$ 
and the quark masses:
\begin{center}
\begin{tabular}{ll}
${\bar{m}}
=(0.007\pm 0.005)$ GeV, & $\alpha _{s}^{n
}=0.65\pm 0.15,$ \\ 
$m_{s}=(0.14\pm 0.03)$ GeV, & $\alpha _{s}^{s}=0.47\pm 0.10,$ \\ 
$m_{c}=(1.28\pm 0.05)$ GeV, & $\alpha _{s}^{c}=0.33\pm 0.03,$ \\ 
$m_{b}=(4.60\pm 0.10)$ GeV, & $\alpha _{s}^{b}=0.27\pm 0.02.$%
\end{tabular}
\end{center}

Within the
model accuracy all results for the well known experimental data 
on the $1^{--}$ and  $0^{-+}$ mesons, which are composed of the quark 
and antiquark with $u$-, $d$-,  $s$-, $c$- or $b$-flavours, are consistent 
with  the presented above model parameters. The evaluated mass spectra of the 
pseudoscalar and vector mesons are displayed in the Table 1.
Thus in the framework of the model the flavor independence of the confinement
potential \cite{doro,mart,quroth} is confirmed as well as the accordance with the asymptotic 
freedom behaviour of the $\alpha _{s}$. 
One can see that   the values of model 
parameters $m_i$ and $\alpha _{s}$  are not in
contradiction with the values of Standard Model constants such as current quark
masses and the strong interaction coupling constant $\alpha _{s}$.  

\begin{center}
4. \underline{Phenomenological mass formulae for isovector $\overline{q}
q^{\prime }-$mesons }
\end{center}
\smallskip

In spite of QCD difficulties which exist in for evaluations of mass spectra
of hadrons consisting of light quarks a number of relations among hadron
masses have been obtained using symmetry or phenomenological considerations,
such as the mass relations \cite{r1} for Regge trajectories \cite{r2} or the
Gell-Mann-Okubo relation \cite{g1,g2}. According to the Regge trajectories
approach a hadron with its spin $J$ and mass $M$ within some errors belongs
to a straight trajectory on the $(M^{2},J)-$plane with a slope $\alpha
^{\prime }$ and a intercept $\alpha _{0}$ 
\begin{equation}
J=\alpha _{0}+\alpha ^{\prime }M^{2}
\label{regge}
\end{equation}
Some hadrons belong to trajectories, so called, daughter trajectories, which
are roughly parallel to the main trajectory and are distinguished with
different values of a radial quantum number $n^{r}$ or $n$, $n=n^{r}+1$ \cite%
{r3,r4,r5}.

Now there is a growing interest in improving the accuracy of existing mass
relations and obtaining these relations in the framework of the QCD or QCD
inspired models. For instance,  in Ref. \cite{petrov} 
on the basis of the renormalization group it has been
shown that the intercept $\alpha _{0}$  cannot be calculated
as a function of the coupling constant.
In Ref. \cite{bad} the expressions for the slope $\alpha^{\prime }$ and 
the intercept $\alpha _{0}$ has been presented in terms of the string tension.
On the base of finite energy QCD sum rules the following mass formulae have been 
obtained  for radially excited $\rho -$mesons 
\cite{k0} and $\pi -$ mesons \cite{k1}:
\begin{eqnarray}
M^{2}(\rho ^{n}) &=&M^{2}(\rho )(2n^{r}+1),  \label{kataev} \\
\qquad M^{2}(\pi ^{n}) &=&M^{2}(\pi ^{\prime })n^{r},
\end{eqnarray}
where $M^{2}(\rho )$ is the mass squared of the $\rho -$meson, $M^{2}(\pi
^{\prime })$ is the mass squared of the first radial excitation of $\pi -$%
meson. In the work \cite{k2} more precise formulae for radially excited
hadrons have been proposed. These formulae in general case can be written as 
\begin{equation}
M_{n}^{2}=M_{0}^{2}+\mu ^{2}(n-1)  \label{anisovich}
\end{equation}
\noindent and describe trajectories on the $(n,M^{2})-$plane with different $%
M_{0}$ values and approximately the same $\mu $ for each trajectory
similarly Chew-Frautschi plots (\ref{regge}) on the $(M^{2},J)-$plane
and Nambu-Veneziano formulae for daughter Regge trajectories.

Interesting relations for orbital and radial exited mesons' mass spectra
have been obtained in the model of chromoelectric tubes  \cite{allen}.
For mesons consisting from a massive quark and massless antiquark
the mass spectrum describes with the formula:
\begin{equation} 
\label{ollrad1}
M^2=\pi \sigma(L+2n+3/2),
\end{equation} 
\noindent where $\sigma$ is the string tension, while for mesons consisting from 
two massless constituents the mass formula is
\begin{equation} 
\label{ollrad2}
M^2=2\pi \sigma(L+2n+3/2).
\end{equation} 

Further on we restrict ourselves to those mesons which have isotopical spin
values equal to $1$, in order to bypass problems \ concerning unknown in
some cases structure and mixing parameters for different mesons with the zero isotopical
spin. We neglect mass splittings within isotopic multiplets, so
systematical errors of the order of $10$ $MeV$ for the phenomenological
scheme considered are without doubt admissible. Moreover, taking into account
the existing experimental errors for meson masses we assume 
$30$$\div$$40MeV$\ as the typical absolute errors of the mass values evaluations, as
in Section 3.

In the framework of relativistic independent quark model a mass formula for $%
\overline{q}q^{\prime }-$mesons has a structure which  differs from
structures of mass formulae in other types of potential models. For
instance, in Refs.\cite{khru} the following mass formula has been considered 
\begin{equation}
M(n^{2S+1}L_{J})=E_{1}(n_{1}^{r},j_{1},c,\varkappa
)+E_{2}(n_{2}^{r},j_{2},c,\varkappa ),  
\label{mnf}
\end{equation}
where the mass terms (mass or energy spectral functions) $E_{i}(n_{i},j_{i},c,%
\varkappa ),i=1,2,$ for a quark and an antiquark are defined as
\begin{equation}
E(n, j, c,\varkappa )=\left\{ 
\begin{array}{c}
c+\varkappa \sqrt{2n^{r}+L+j-1/2},\quad L+j-1/2=2k,\quad k=0,1,... \\ 
\varkappa \sqrt{2n_{r}+L+j-1/2},\quad L+j-1/2=2k+1,\quad k=0,1,...%
\end{array}
\right.  
\label{enf}
\end{equation}

In order to exclude superfluous meson states, the following selection rules
for \textbf{\ }$\overline{q}q^{\prime }-$ me\-sons with given $J^{PC}$ values,
quark masses $m_{1}$ and $m_{2}$ and quantum numbers $j_{1}$ and $j_{2}$ are
used 
\begin{equation}
\begin{array}{l}
j_{1}=j_{2}=J+1/2,\quad if\quad J=L+S, \\ 
j_{1}=j_{2}+1=J+3/2,\quad if\quad J\neq L+S,\quad m_{1}\leq m_{2,}%
\end{array}%
\label{jnf}
\end{equation}
while the radial quantum numbers $n_{1}^{r}=n_{2}^{r}=n-1$ for $%
n{}^{2S+1}L_{J}$ - state. The meson parity $P=(-1)^{L+1}$ and the eigenvalue 
$C$ of charge conjugation for the neutral meson of $q\bar{q}$ type is equal
to $(-1)^{L+S}$, where the total spin $S=0$ or $1$.

Functions $E_{i}(n_{i}^{r},j_{i},c,\varkappa )$ , $i=1,2$, give the
relativistic effective energies of the quark and antiquark moving in the
mean field inside the meson, and include also an energy of the mean field
and possible nonpotential corrections, which cannot be evaluated within the
mean field approximation and  according to the formula (\ref{mava}) should 
be taken into account with the help of the constants $E_{0m}$.
 Note that nonpotential corrections are the most important for
mass spectra of light mesons \cite{fili,diek}. In general case
the main part of mass term $E_{i}(n_{i}^{r},j_{i},c,\varkappa )$, 
as it has been described in Section 2, are determined from the solution of 
the Dirac equation. However, for the superlight mesons, which consist
of $u-$ and $d-$ quarks and antiquarks only, the phenomenological energy 
spectral functions $E_{i}(n_{i},j_{i},c,\varkappa )$ in the form (\ref{enf}) 
are suitable within the  assumed accuracy. When evaluating masses of unknown 
excited meson states  we use the formulae
written above together with the values of two parameters $c$ and 
$\varkappa $, which have been obtained by fitting the mass values of 
experimentally detected meson states. The $c$ and $\varkappa $ values 
obtained by this manner are $c=69$ $MeV,$ $\varkappa =382\pm 4$ $MeV$ 
\cite{kh1}. The results obtained with $c=69$ $MeV,$ $\varkappa =385$ $MeV$ 
are shown in the Table 2, where we present the mass values of orbital 
excitations of superlight meson states up to $L=4$.

One can obtain with the help of the formulae (\ref{mnf}), (\ref{enf}) and 
the selection rules (\ref{jnf}) a great number of mass relations, which are 
fulfilled  within the systematical
errors of the phenomenological scheme considered. For instance, in the case
of orbitally excitated vector mesons formulae (\ref{mnf}) and (\ref{enf}) 
give the $\rho -$trajectory, while for radially excitated vector mesons they 
bring in the formula (\ref{kataev}). In general case in the framework of 
this approach instead of the meson trajectories in $(M^{2},J)-$ and 
$\ (n, M^{2})-$planes we have in $(L,n,M)-$space the following four series 
or trajectories  for $n^{2S+1}L_{J}-$ meson states.

There is a $(\pi \rho )-$ series for $n^{3}L_{L-1}-$meson states with $%
P=(-1)^{L+1}$, $C=(-1)^{L+1}$. Mass values of the members of the $(\pi \rho
)-$ series are determined by the formula: 
\begin{equation}
M_{n,L}^{\pi \rho}=c+\varkappa \sqrt{2(n^{r}+L)-1}+\varkappa \sqrt{2(n^{r}+L)%
}
\end{equation}

There is a $\pi -$ series for $n^{1}L_{L}-$meson states with $P=(-1)^{L+1}$, 
$C=(-1)^{L}$. Mass values of the members of the $\pi -$ series are
determined by the formula: 
\begin{equation}
M_{n,L}^{\pi}=2c+2\varkappa \sqrt{2(n^{r}+L)}
\end{equation}

There is a $(\rho \pi )-$ series for $n^{3}L_{L}-$meson states with $%
P=(-1)^{L+1}$, $C=(-1)^{L+1}$. Mass values of the members of the $(\rho \pi
)-$ series are determined by the formula: 
\begin{equation}
M_{n,L}^{\rho\pi}=c+\varkappa \sqrt{2(n^{r}+L)}+\varkappa \sqrt{2(n^{r}+L)+1}
\end{equation}

There is a $\rho -$ series for $n^{3}L_{L+1}-$meson states with $%
P=(-1)^{L+1} $, $C=(-1)^{L+1}$. Mass values of the members of $\rho -$
series are determined by the formula: 
\begin{equation}
M_{n,L}^{\rho}=2\varkappa \sqrt{2(n^{r}+L)+1}
\label{rom}
\end{equation}

It should be noted that only two phenomenological parameters enter in 
these formulae, which are in actual fact equal to half the pion mass and
half the $\rho -$meson mass. With the help of these formulae for the series 
presented many mass
relations which are not dependent on intrinsic quantum numbers $L$ and $n$
easily can be obtained. For instance, mass values of members of $\pi -$
series, $(\rho \pi )-$ series and $\rho -$ series with the same $L$ obey the
following mass relation: 
\begin{equation}
2M_{n,L}^{\rho\pi}(n^{3}L_{L})=M_{n,L}^{%
\rho}(n^{3}L_{L+1}) +  M_{n,L}^{\pi}(n^{1}L_{L})\label{mnl}
\end{equation}

\begin{center}
5. \underline{Link of  model of quasi-independent quarks with the
model of constituent quarks} 
\end{center}
\smallskip

As indicated above in order to improve the accuracy of calculations for meson 
mass spectra in the framework of quasi-independent quarks model
an account of residual interactions'
contributions jointly with  different values for mean field
energy is needed. We find  in Sec. 3 the
values of mean field energy   in combination with
magnitudes of spin-spin interaction for quarks and antiquarks with
different flavours inside the pseudoscalar and vector mesons.
Further on  we propose that these
values are not simply some phenomenological constants, but
represent in a rough approximation equidistant levels for mean
field energy, which do not depend on quark flavours. 
 We consider ordinary
(non exotic) hadrons and relate the model of quasi-independent
quarks to the constituent quark model.  It will be obtained that
energies of fermionic constituents $E_i$ (quark mass terms) for meson
ground states are equal to constituent quarks' masses within
uncertainties of model of quasi-independent quarks. Some
complications arise when one treats exited states. In the model of
quasi-independent quarks $E_i$ remains an exited energy of i-th
constituent, while in the constituent quark model besides
constituent quarks' masses an additional contribution representing
energy of excitation must be taken into consideration.
  For the pseudoscalar and vector mesons the spin-spin interaction
may significantly account for a residual interaction in the ground
states of $(\bar qq')$ system within the accuracy of the model and
mesons' masses can be evaluated by means of the  formula (\ref{me})
presented in Sec.3. It should be noted that in the framework
of the constituent quark model one can find the similar formula
\cite{zeldsa,sa}:
\begin{equation}
 M_m =  m_0 +  m_1 + m_2 + (\mathbf{s}_{1}\mathbf{s}_{2})v_{ss},
\label{fueqq}
\end{equation}
where $m_1$, $m_2$ are  masses of constituent quarks, $m_0$ is some additional
phenomenological contribution.  So we relate constituents'
energies $E_i , i = 1, 2,$ with constituent quarks' masses $m_i, i = 1,
2$. In what follows this assumption will be supported by numerical
fit for meson masses. Formulae (\ref{me}) and (\ref{fueqq}) which 
correspond each other may be named the Zeldovich-Sakharov formulae.
  Note that constituent quarks can be applied
successfully not only in hadron spectroscopy, but  for description
of effects in hadron collisions as well  (see, e.g. \cite{tt}).

  At present due to experimental uncertainties and mixing effects
there is no only one solution for the set of parameters entering
in Eq.(\ref{me}). For instance, there is a well known difficulty related
to the mass value of the $\pi-$meson. Below we give an admissible fit
with minimal number of parameters for the mass values of
pseudoscalar and vector mesons in the ground states, which can be considered as
definite $(\bar q q')$ systems involving $u-, d-, s-, c-, b-$quarks and
antiquarks.
  To do this, we use in an essential manner the results obtained in Sec.3.
As indicated above in
 spite of the fact that the quark and antiquark energies $E_1$
and $E_2$ bring in a main contributions to meson masses in order to
reduce relative uncertainties of evaluations up to $10^{-2}$ level for
all mesons the additional contributions due to the $E_0$ and $V_{ss}$ are
needed. Along similar lines $E_0$ and $V_{ss}$ contributions were
considered in certain of approaches \cite{fili,zeldsa,sa}, however
some ambiguities occur when $E_0$ and $V_{ss}$ are evaluated. When
different fits for meson masses had been performed the case was
adopted, such that the $V_{ss}$ value for $u-,d-$quarks and antiquarks
was equal to $100 MeV$ and the $E_0$ value for $\rho-$meson was zero. In this
case the $E_0$ value for $\pi -$meson is double that the $E_0$ value for $K$
meson: $E_0^{\pi} = 2E_0^K$, where $E_0$ values are marked at the top with
mesons' names. The nonzero values of $E_0$ and $V_{ss}$ given in MeV units
for different pseudoscalar and vector mesons are listed below.
\begin{eqnarray}
  E_0^K = -118,  E_0^{J/\psi} = -125,  E_0^{\eta_c} = -230, 
E_0^{\eta_b} = -600,  \nonumber\\
E_0^{\Upsilon} = -450,  E_0^{D_s} = -112,   E_0^{B_c} = -270, V_{ss}^{K^*} = 70, 
V_{ss}^{\phi} = 50, \\
V_{ss}^{D^*} = 27,  V_{ss}^{D_s^*} = 5,   V_{ss}^{J/\psi} = 3, V_{ss}^{B^*} = 3, 
V_{ss}^{B_s^*} = 2, V_{ss}^{D_c^*} = 1.3,V_{ss}^{\Upsilon}
= 1\nonumber
\end{eqnarray}
  The uncertainties of determination of the $E_0$ and $V_{ss}$ values are
of the order of several MeV. As is seen in the list of the $E_0$
values above all nonzero $E_0$ values for ground states of
pseudoscalar and vector mesons for all quark flavours are
approximately multiples to the $E_0^K$. Thus we can accept that the $E_0$
values  are varied according to the rule:
\begin{equation}
                        E_0^n = -708 + n E_0^K,
\end{equation}                                
where n is a negative integer. To give also the values of
constituents' energies $E_i$ for $u-, d-, s-, c-, b-$quarks and
antiquarks (for  two super light $u-$ and $d-$quarks and antiquarks
the mean value $E_N$ is presented):
\begin{equation}
  E_N = 337\pm 3,  E_S = 485\pm 8, E_C = 1610\pm 15, E_B = 4952\pm
20  
\label{const1}
\end{equation}
Notice that within the model uncertainties the constituents'
energies for light quarks and antiquarks agree with the
constituent quark masses obtained in Ref.\cite{sca}. Moreover if one take 
into account the difference between masses of U- and D-quarks obtained in this 
work, which is equal to 4 MeV, the masses of   U- and D-quarks get the following values:
\begin{equation}
  E_U = 335\pm 2, \quad E_D = 339\pm 2 
\label{const2}
\end{equation}
Masses of constituent quarks  (\ref{const1}) and (\ref{const2})  can be
used for the derivation of such important parameters of Standard Model as 
quark mixing angles in the Fritzsch-Scadron-Delbourgo-Rupp approach 
\cite{fri,sca,gakhse}.

  The model of quasi-independent quarks supplemented by the
proposition about the equidistant discrete levels of mean field
may be named the extended model of quasi-independent quarks. In
the framework of this model it is possible to evaluate the
pseudoscalar and vector mesons' masses with uncertainties of order
$10^{-2}$. From the
results presented it may be inferred that the extended model of
quasi-independent quarks is a workable generalization of the constituent
quarks model. 

\begin{center}
6.\underline{ Conclusions and discussion} 
\end{center}
\smallskip

It is highly plausible that the formation of
constituent quarks take place in the nonperturbative region with
the characteristic scales, which can be evaluated with the
universal coefficient of a slope of a linear rising potential $\sigma =
0.20\pm 0.01 GeV^2$. The scales' values with dimensions of mass and
length are equal to $\mu_C = 0.45\pm0.02 GeV$, $\lambda_C = 0.44\pm0.02 Fm$.
 As this
takes place, $\mu_C$ defines typical magnitudes of transversal impulses
$<p_T>$ for quarks-partons inside hadrons, while a radius of a
perturbative region surrounded a current quark is equal to $r_C =
\lambda_C/2 = 0.22\pm0.01 Fm$. The region $r > r_C$  is most likely to be the
region of a formation for a constituent quark due to
nonperturbative interactions.  If one take into account typical sizes
of hadrons, then a radius of a constituent quark is $0.25 \div 0.35 Fm$.
The absolute value of  $E_0^K$, which
determinates the characteristic scale for mean field energies in
the framework of the extended model of quasi-independent quarks
can also be expressed in terms of $\mu_C$ as $\mu_C/4$.

There is a widespread opinion, that the 
confinement of quarks and gluons can be established rigorously in the QCD 
framework. Is this is the case, a exact dependence should take place, which 
relates $\Lambda_{QCD}$ to the characteristic parameter of strong interactions
 in the  nonperturbative region -- the  string tension $\sigma$. However, 
in the case if the confinement cannot be demostrated in the QCD framework, 
there has to add the number of strong interaction parameters and to introduce,
 at the least, one more parameter -- the string tension. From this standpoint 
of some interest is investigations concerning with the generalization of the 
space-time Poincare symmetry of the QCD in the confinement region, for
instance, to the 
inhomogeneous pseudounitary symmetry  IU(3,1) \cite{kh2}, and in this case 
the universality of the confinement potential verified in the framework of the 
relativistic model of quasi independent quarks have appresiable significance.

In this report we considered the relativistic hadron model with 
quasi-independent  quarks and presented the results of the evaluations of 
mass spectra of radially and orbitally exited meson states. It is important
to compare evaluated meson masses with experimentally determined ones 
\cite{eidel}. As it is seen from Tables 1 and 2 there is an agreement between 
the evaluated and experimentally determined values on the level
of $1-2\sigma$. For instance, let us 
compare with the data \cite{eidel} the mass relation (\ref{mnl}), which
must obey for the mesons with $L\neq 0$ and $S=1$. At present this relation
can be checked only for the mesons with $L=1.$ After the substitution the
experimental values of the masses for the $a_{1}(1260)$, $b_{1}(1235)$ and $%
a_{2}(1320)$ mesons, we obtain, that this relation is fulfilled with account
of experimental uncertanties. Moreover, the mass formulae presented above
permit to explain the degeneracy with respect to mass values for different $%
J^{PC}$ mesons. For example, the mass formulae (\ref{rom}) give $M(\rho^{\prime%
\prime})=M(\rho_3)$. If one use the data, then $1720\pm20 MeV=1688.8\pm2.1
MeV$. In the high mass value region the degeneracy of such kind will be
increased, as it is seen in the $\sim 2315 MeV$ region, where the 
mass values for $0^{-+}$, $%
1^{--}$, $4^{-+}$, $4^{--}$ and $5^{--}$ mesons are predicted.

As it follows from the Table 2, the mass value calculated for 
orbital excitation of the vector $1^{--}$ meson is equal to $1506 MeV$. 
So we expect that in the mass region
between \ 1.3 GeV and 1.6 GeV two standard $\overline{q}q^{\prime }-$meson
resonances with $J^{PC}=1^{--}$ and $I=1$ exist, namely the first radial and
orbital excitations mixing each another. However,
it is possible that the more complicated situation should be considered
when in the mass region $1.2 \div $ $1.8$ $GeV$ 
the mixing of the standard $1^{--}$
mesons and the vector non $\overline{q}q^{\prime }-$mesons take place as well.
We considered above  only possible mixing between the first radial 
and orbital excitated standard $1^{--}$ mesons.
The most complicated situation may arise, which is not disscussed here, when
the mixing between two radial and orbital 
exitations or with  vector cryptoexotic states can occur. These cases 
demand further investigation. For instance, in the framework of the
well-known potential model \cite{godf} the mass values of $2$ $^{3}S_{1}$, $%
1^{3}D_{1}$ and $3$ $^{3}S_{1}$ states lie considerably higher (at $1.45$,
$1.66$, $2.00$ $GeV$, correspondingly) than the predictions 
considered. Moreover, in Ref. \cite{faust} the mass value $1486 MeV$ has
been evaluated the first radial excitation of $\rho-$meson. In Ref.\cite{piv}
the mass values for the radial exitations of $\rho-$meson, which are equal to
1.4, 1.8, 2.13 Gev, have been obtained.  Some difficulties concerning with the 
identification of the excited states of $\rho-$meson are available in
Ref.\cite{acko}.

Another importain problem is the interpretation of mesons discovered in
scalar channel. The authors of Ref.\cite{k2} found that in the region
$\sim 1 GeV$ there are $P-$wave 
\textbf{\ }$\overline{q}q^{\prime }-$mesons. The results
of our evaluations (Table 2)  support the existence of the $P-$wave 
\textbf{\ }$\overline{q}q^{\prime }-$mesons with masses $\sim 980$ $MeV$.
The complementary reasoning in favour of the existence of
the $P-$wave \textbf{\ }$\overline{q}q^{\prime }-$mesons with masses $\sim
980$ $MeV$ is the coincidence of the evaluated mass value for the first
radial excitation of $a_{0}(980)$ meson  with the mass value of the 
$a_{0}(1450)$ meson \cite{eidel}. This meson has been predicted 
for the first time in Ref. \cite{gor} in the framework of finite energy
QCD sum rules.
Although the mass values of the $a_{0}(980)$ and\ $\
f_{0}(980)$ mesons show the 
possible ideal mixing between them the problems with the
intensities of different decay modes exist (in particular, the$\ K\overline{K}
$ decay mode enhancement). The  possible explanation of these facts
now comprises the four-quark nature of the $a_{0}(980)$ and\ $\ f_{0}(980)$
mesons \cite{jaffe,acha}.
However, there is an accidental degeneracy of scalar mesons mass values with 
the value of $%
\overline{K}K-$threshold, which complicates the mechanism of an extraction 
of $%
a_{0}(980)$ and\ $\ f_{0}(980)$ decays characteristics \cite{baru}. Moreover,
it is plausible that the mixing of these mesons with 
the $\overline{K}K-$molecule arises \cite{kal}. 
Taking into account the existing difficulties with identification of 
light mesons in the $1\div2$ GeV mass region we see that 
allowing for the mixing between  two or three mesons with different 
structures but in some cases with about the same masses
is admissible for an adequate data description  in
both scalar  and   vector channels.

\smallskip

     The  author  is grateful to Yu.V.  Gaponov,  A.E.
Dorokhov,  A.L. Kataev, A.V. Leonidov  V.I.
Savrin, S.V. Semenov,A.M. Snigirev  and  V.E. Troizky for useful discussions
and the organizers of the PNPI Winter School for the hospitality. The
work  was  partially supported by the  grant   26  on  fundamental
researches of the RRC "Kurchatov Institute" in 2007 year.

\vspace{0.5cm}

\bigskip 
\newpage
\hspace*{0.5cm}{\small Table 1. }

\hspace*{0.5cm}{\small\parbox[t]{13.9cm}{Evaluated masses 
 of the $1^{--}$ and $0^{-+}$ mesons 
in comparison with the  data from Ref. \cite{eidel}.}}

\begin{center}
\begin{tabular}{|c|c|c|c|c|c|}
\hline
Meson & $M_{n}^{exp[1]}[MeV]$ & $M_{n}^{th}[MeV]$ & 
Meson & $M_{n}^{exp[1]}[MeV]$ & $M_{n}^{th}[MeV]$ \\ 
\hline
$\rho $ & 775.5$\pm $0.4 & 740 & $B$ & 5279$\pm $0.8 & 5250  \\ 
$\rho ^{\prime }$ & 1459$\pm $11 & 1455 & $B_{s}$ & 5367.5$\pm $1.8 & 5370\\ 
$\rho ^{\prime \prime }$ & 1720$\pm $20 & 1730 & $B_{c}$ & 6286$\pm $5 & 6300  \\ 
$\phi $ & 1019.460$\pm $0.019 & 1010 & $\eta _{b}$ & 9300$\pm $40 & 9330 \\ 
$\phi ^{\prime }$ & 1680$\pm $20 & 1650 &  $\pi ^{\prime }$ & 1300$\pm $100 & 1290\\ 
$\phi ^{\prime \prime }$ & - & 2050 & $K^{\prime }$ & - & 1400\\ 
$J/\psi $ & 3096.916$\pm $0.011 & 3060 & $D^{\prime }$ & - & 2450\\ 
$\psi ^{\prime }$ & 3686.093$\pm $0.034 & 3650 & $D_{s}^{\prime }$ & - & 2560\\ 
$\psi ^{\prime \prime }$ & 4039$\pm $01 & 4070 & $\eta _{c}^{\prime }$ & 3638$\pm $4 & 
3600\\ 
$\psi ^{\prime \prime \prime }$ & 4421$\pm $4 & 4390 &$B^{\prime }$ & - & 5650 \\ 
$\Upsilon $ & 9460.30$\pm $0.26 & 9470 & $B_{s}^{\prime }$ & - & 5750\\ 
$\Upsilon ^{\prime }$ & 10023.26$\pm $0.31 & 9990 &  $B_{c}^{\prime }$ & - & 6800 \\ 
$\Upsilon ^{\prime \prime }$ & 10355.2$\pm $0.5 & 10325 & $\eta _{b}^{\prime }$ & - & 9960 \\ 
$\Upsilon ^{\prime \prime \prime }$ & 10579.4$\pm $1.2 & 10550 & $\pi
^{\prime \prime }$ & 1812$\pm $14 & 1810 \\ 
$\Upsilon ^{5S}$ & 10865$\pm $8 & 10830 & $K^{\prime \prime }$ & -
& 1850 \\ 
$\Upsilon ^{6S}$ & 11019$\pm $8 & 10985 & $D^{\prime \prime
} $ & - & 2880 \\ 
$\pi $ & 138$\pm $3.1 & 120 & $%
D_{s}^{\prime \prime }$ & - & 2950 \\ 
$K$ & 495.65$\pm $0.02 & 500 & $\eta _{c}^{\prime \prime }$ & - & 3970 \\ 
$D$ & 1867.7$\pm $0.4 & 1850& $B^{\prime \prime }$
& - & 6070 \\ 
$D_{s}$ & 1968.2$\pm $0.5 & 1990& $%
B_{s}^{\prime \prime }$ & - & 6130 \\ 
$\eta _{c}$ & 2980.4$\pm $1.2 & 2990& $%
B_{c}^{\prime \prime }$ & - & 7150  \\
\hline
\end{tabular}

\bigskip
\end{center}

\medskip

\bigskip

\bigskip {\small Table 2. }

{\small 
}

{\small 
\parbox[t]{16.cm}{Evaluated masses in MeV for the ground states  and the
orbital excitations of the   $\bar qq'-$ mesons in
comparison with the  data from Ref.\cite{eidel}. }}

\medskip

\begin{tabular}{cccccccc}
\hline
Meson & $J^{PC}$ & $M^{exp}(MeV)^{[17]}$ & $M^{ev}(MeV)$ & Meson & $J^{PC}$
& $M^{exp}(MeV)^{[17]}$ & $M^{ev}(MeV)$ \\ \hline
$\pi $ & 0$^{-+}$ & 138.039$\pm $0.004 & 138$\pm $30 & $\rho _{3}$ & 3$^{--}$ & 
1688.8$\pm $2.1 & 1722$\pm $30 \\ 
$\rho $ & 1$^{--}$ & 775.5$\pm $0.4 & 770$\pm $30 & $a_{2}^{3}$ & 2$^{++}$ & 
- & 1873$\pm $30 \\ 
$a_{0}$ & 0$^{++}$ & 984.7$\pm $1.2 & 998$\pm $30 & $b_{3}^{3}$ & 3$^{+-}$ & 
- & 2024$\pm $30 \\ 
$b_{1}$ & 1$^{+-}$ & 1229.5$\pm $3.2 & 1227$\pm $30 & $a_{3}^{3}$ & 3$^{++}$
& - & 2031$\pm $30 \\ 
$a_{1}$ & 1$^{++}$ & 1230$\pm $40 & 1280$\pm $30 & $a_{4}$ & 4$^{++}$ & 2001$%
\pm $10 & 2037$\pm $30 \\ 
$a_{2}$ & 2$^{++}$ & 1318.3$\pm $0.6 & 1334$\pm $30 & $a_{3}^{4}$ & 3$^{--}$
& - & 2177$\pm $30 \\ 
$a_{1}^{2}$ & 1$^{--}$ & - & 1506$\pm $30 & $b_{4}^{4}$ & 4$^{-+}$ & - & 2316%
$\pm $30 \\ 
$\pi _{2}$ & 2$^{-+}$ & 1672.4$\pm $3.2 & 1678$\pm $30 & $a_{4}^{4}$ & 4$%
^{--}$ & - & 2313$\pm $30 \\ 
$a_{2}^{2}$ & 2$^{--}$ & - & 1700$\pm $30 & $a_{5}^{4}$ & 5$^{--}$ & - & 2310%
$\pm $30 \\ \hline
\end{tabular}
\end{document}